\input amstex
\documentstyle{amsppt}

\mag=\magstep1
\vsize=21.6truecm
\hsize=16truecm
\NoBlackBoxes
\leftheadtext{Yunbo Zeng and Wen-Xiu Ma}
\rightheadtext{Families of Quasi-Bi-Hamiltonian systems and
Separability}

\topmatter
 \title Families of Quasi-Bi-Hamiltonian Systems and Separability
\endtitle
\author
Yunbo Zeng$^{*}$ \footnote { E-mail:yzeng\@tsinghua.edu.cn}
and Wen-Xiu Ma$^{\dagger}$\footnote {E-mail:mawx\@cityu.edu.hk}
\endauthor
\affil
*Department of Mathematical Sciences,
Tsinghua University\\ Beijing 100084, China\\
$\dagger$ Department of Mathematics,
City University of Hong Kong\\
Kowloon, Hong Kong, China\\
\endaffil
\endtopmatter
\def\p{{\partial}}
\def\la{{\lambda}}
\def\La{{\Lambda}}

\TagsOnRight
{\bf Abstract.} It is shown how to construct an infinite number of
families of quasi-bi-Hamiltonian (QBH) systems by means of the
constrained flows of soliton equations. Three explicit QBH
structures are presented for the first three families of the
constrained flows. The Nijenhuis coordinates
defined by the Nijenhuis tensor for the corresponding families of
QBH systems are proved to be exactly the same as the separated
variables introduced by mean of the Lax matrices for the constrained
flows.


\ \par

\newpage

\subhead {I. Introduction}\endsubhead\par
\ \par

As is known, some integrable systems possess bi-Hamiltonian structure.
We recall some known results. Let $M$ be a differential manifold, $TM$
and $T^*M$ its tangent and cotangent bundle, and $\theta_0$  and
$\theta_1:T^*M \rightarrow TM $ two compatible Poisson tensors on
$M$ [1]. A vector field $X$ is said to be bi-Hamiltonian (BH) with
respect to
$\theta_0$ and $ \theta_1$, if two smooth functions, $H, F\in
C^{\infty}(M)$,
 exist such that
$$X=\theta_0dH=\theta_1dF \tag 1.1$$
where $dF$ denotes the differential of $F$ (gradient $\bigtriangledown
F$
for finite system and variation $\delta F$ for field system).
If $\theta_0$ is invertible, the tensor $\Phi=\theta_1\theta_0^{-1}$ is
a
Nijenhuis tensor or hereditary operator. The operator $\Phi$ maps a
given
BH vector field into another BH vector field. Hence having a Nijenhuis
tensor, one can construct a hierarchy of Hamiltonian symmetries,
and a related hierarchy of integrals of motion for the underlying
system.
The BH structure (1.1) ensures that the resulting integrals of motion
are
pairwise in involution with respect to both Poisson brackets. Thus the
BH
structure of a given system  is important for its integrability.\par
Unfortunately, for a majority of the BH finite-dimensional systems,
none
of the $\theta_0$  and  $\theta_1$ is invertible. In fact, all the
known
BH finite-dimensional systems arising from the constrained flows or
stationary flows of soliton equations usually exist in an extended
phase space and both $\theta_0$  and  $\theta_1$ are degenerated (see,
for example, [2-8]). In their natural phase space these systems may
satisfy a weaker condition than the BH one. The notion of a
quasi-bi-Hamiltonian (QBH) system was introduced [9,10]. According
to [10], for $dimM=2n$, a vector field, $X$, is said
to be a QBH vector field with respect to Poisson tensors, $\theta_0$
and  $\theta_1$, if there exist three smooth functions $H, F, \rho,$
such that
$$X=\theta_0\bigtriangledown H=\frac 1{\rho}\theta_1\bigtriangledown
F \tag 1.2$$
where two Poisson tensors $\theta_0$ and $ \theta_1$ are compatible
and nondegenerated (invertible). The function $\rho$ is called an
integrating factor. On a $2n$-dimensional symplectic manifold $M$,
let $(\pmb{q}=(q_1,...,q_n),  \pmb{p}=(p_1,...,p_n))$ be a set of
canonical coordinates and $\theta_0$ the canonical Poisson matrix
$\theta_0=\left(\matrix 0&I\\-I&0\endmatrix\right)$
($I$ denoting the $n\times n$ identity matrix). As $\theta_0$ and
$ \theta_1$ are compatible and invertible, the Nijenhuis tensor
$\Phi=\theta_1\theta_0^{-1}$ is maximal, i.e. it has $n$ distinct
eigenvalues $\pmb{\mu} =(\mu_1,...,\mu_n)$. As is known [11], in
a neighborhood of a regular point, where the eigenvalues $\pmb {\mu}$
are distinct, one can construct a canonical transformation
$(\pmb{q}, \pmb{p}) \mapsto (\pmb{\mu}, \pmb{\nu})((\pmb{\mu},
\pmb{\nu})$
referred to as the Nijenhuis coordinates) such that $ \theta_1$ and
$\Phi$
take the Darboux form
$$\theta_1=\left(\matrix 0&\La_1\\-\La_1&0\endmatrix\right), \qquad
\Phi=\left(\matrix \La_1&0\\0&\La_1\endmatrix\right), \qquad
\La_1=diag (\mu_1,...,\mu_n). \tag 1.3$$
A QBH vector field is said to be Pfaffian [10] if, in the Nijenhuis
coordinates, an integrating factor $\rho$ in equation (1.2) is the
product of the eigenvalues of $\Phi$, i.e.
$$\rho=\prod_{i=1}^n \mu_i. \tag 1.4$$\par
In the Pfaffian case, the general solutions, $H$ and $F$, of equation
(1.2) are obtained and the Hamilton-Jacobi equation for $H$ is shown
to be separable by verifying the Levi-Civita conditions [12]. Some
relationship between BH and QBH structure is discussed in [13]. Several
QBH systems are presented  [9,10,12-14]. It is in general quite
difficult
to directly construct a BH or QBH structure for a given integrable
Hamiltonian vector field.
In recent years much work has been devoted to the constrained flows
of soliton equations (see, for example, [2-8,15-24]).
One of the aims of this paper is to show how to construct an infinite
number of families of QBH systems from the constrained flows of soliton
equations. We have presented some families of the constrained flows in
order to study the dynamical $r$-matrices in [24]. We now describe the
explicit QBH structures for these families of the constrained flows.\par

The Lax representation for the constrained flows of soliton equations
can always be deduced from the adjoint representation of the Lax pair
for soliton equations [16,17]. There is an effective way for the
separation of variables for some finite-dimensional integrable
Hamiltonian systems with some kind of Lax matrices [25,26]. The
separated variables for some constrained flows can be introduced
and the Jacobi inversion problems for the constrained flows can
be established by means of the Lax representation [27,28]. We are
interested in the relationship between the two methods for the
separability mentioned above. Another main aim of this paper is to
prove that the Nijenhuis coordinates for the underlying families of
QBH systems are usually the same as the separated variables introduced
by the Lax matrices.\par
The paper is organized as follows.\par
In section 2 we present a new QBH system. We directly construct the
second compatible Poisson tensor by using a map relating this system
to its modified version, and prove that the Nijenhuis coordinates for
this system is equivalent to the separated variables defined by Lax
matrix. We make some comparison of the two methods for separability.
In section 3 and section 4, by using the constrained flows associated
with the polynomial second order spectral problems and the higher-order
symmetry constraints, we propose a way to construct an infinite number
of families of QBH systems. The explicit QBH structures of the first
two families of constrained flows are given. The equivalence of the
Nijenhuis coordinates and the separated variables is proved. In
section 5 we point out that the two compatible Poisson tensors
$\theta_0, \theta _1$ and the integrating factor $\rho$ given
by the QBH structure (2.28) and (2.29a) are just that for the
third family of OBH systems. Also some conclusions and a conjucture
are given.
\par
\par
\subhead {II.  New QBH system}\endsubhead\par
\ \par
In this section we present a new QBH system. By using a map relating
this system to its modified version, the second compatible Poisson
tensor is obtained from the image of the Poisson tensor for the
modified version under the map. We use this system to illustrate
how to prove the equivalence of the Nijenhuis coordinates and the
separated variables introduced by the Lax matrix.\par
\subhead {A. New finite-dimensional integrable Hamiltonian system}
\endsubhead\par
\ \par
For Jaulent--Miodek (JM) spectral problem [29]
$$\psi_x=U(u,\la)\psi,\quad
 U(u, \lambda)
=\left( \matrix 0&1\\\la^2-u_1\la-u_0
&0\endmatrix\right),\quad
\psi=\binom {\psi_{1}}{\psi_{2}},\quad u=\binom {u_1}{u_0},\tag 2.1$$
its adjoint representation is defined by [30]
$$V_x=[U, V]\equiv UV-VU,\tag 2.2$$
where $V$ is taken as
$$V=\sum_{i=0}^{\infty}V_i\la^{-i}, \qquad V_i
=\left( \matrix a_i&b_i\\c_i&-a_i\endmatrix\right).\tag 2.3$$
Then equation (2.2) and (2.3) yields
$$a_{0}=a_1=a_{2}=b_{0}=b_{1}=0,\quad b_{2}=1,\quad
b_{3}=\frac{1}{2}u_{1},$$
$$a_{3}=-\frac{1}{4}u_{1,x},
\quad c_{0}=1,\quad c_{1}=-\frac{1}{2}u_{1},\hdots,$$
and in general
$$\binom{b_{k+2}}{ b_{k+1
}}=L \binom{b_{k+1}}{ b_{k}}, \qquad k=1,2,...\tag 2.4a$$
$$ a_k=-\frac 12b_{k,x},\qquad
c_{k}=a_{k,x}-u_0b_k-u_{1}b_{k+1}+b_{k+2},
\quad k=1,2,\cdots,\tag 2.4b$$
where
$$L=\left( \matrix u_1-\frac 12D^{-1}u_{1,x}&$$\frac{1}{4}D^2+u_0
-\frac 12D^{-1}u_{0,x}
\\1&0\endmatrix\right),\qquad D=\frac {\p}{\p x},\quad
DD^{-1}=D^{-1}D=1.$$
The Jaulent-Miodek hierarchy associated with (2.1) can be written as an
infinite-dimensional
Hamiltonian system
$$u_{t_n}={\binom {u_{1}}{u_0}}_{t_n}
=J\binom {b_{n+2}}{ b_{n+1}}
=J\frac {\delta H_{n}}{\delta u},\qquad n=1,2,\hdots, \tag 2.5$$
where the Hamiltonian $H_n$ and the Hamiltonian operator $J$ are given
by
$$J=\left( \matrix 0&2D\\2D&-u_{1x}-2u_1D\endmatrix\right),\qquad
H_{n}=\frac{1}{n}(2b_{n+3}-u_{1}b_{n+2}).$$
Under zero boundary condition we have
$$\frac {\delta \lambda}{\delta u}=\binom {\la
\psi_{1}^{2}}{\psi_{1}^{2}},
\qquad
L\frac {\delta \lambda}{\delta u}=\la\frac {\delta \lambda}{\delta
u}.\tag 2.6$$
\par
The constrained flow of (2.5) consists of the equations obtained from
 the spectral problem  (2.1) for $N$
distinct $\lambda_j$ and the restriction of the variational
derivatives for conserved
quantities $H_{l}$ (for any fixed $l$) and $\lambda_{j}$ [15-17]:
$$ \Psi_{1,x}=\Psi_{2},\qquad
\Psi_{2,x}=\La^2\Psi_{1}-u_1\La\Psi_1-u_0\Psi_1,\tag 2.7a$$
$$\frac {\delta H_{l}}{\delta u}-\frac 12
\sum_{j=1}^{N}\frac {\delta \lambda_{j}}{\delta u}
=\binom {b_{l+2}}{ b_{l+1}}-\frac 12\binom {
<\La\Psi_1,\Psi_1>}{ <\Psi_1,\Psi_1>}=0,\tag 2.7b$$
which has been recognized as a symmetry constraint [18-20].
Hereafter we denote the inner product in $\text{\bf R}^N$ by $<.,.>$ and

$\Psi_i=(\psi_{i1},\cdots,\psi_{iN})^{T},$ $i=1,2,  \Lambda=diag
(\lambda_1,\cdots,\lambda_N).$\par

For $l=4$, we have
$$H_4=\frac 7{128}u_1^5+\frac 5{16}u_{1}^3u_{0}-\frac 5{32}
u_{1x}^2u_{1}
+\frac 3{8}u_{0}^2u_{1}-\frac 1{8}u_{1x}u_{0x}.\tag 2.8$$
By introducing the Jacobi-Ostrogradsky coordinates
$$q_1=u_1,\qquad q_2=u_0,$$
$$p_1=\frac {\delta H_4}{\delta u_{1x}}=-\frac 5{16}u_{1}u_{1x}
-\frac 1{8}u_{0x},
\qquad p_2=\frac {\delta H_4}{\delta u_{0x}}=-\frac 1{8}u_{1x},\tag
2.9$$
the equations (2.7) for $l=4$ are transformed into a finite-dimensional
Hamiltonian system (FDHS)
$$ \Psi_{1x}=\frac {\p F_1}{\p \Psi_2}=\Psi_2,\quad
q_{1x}=\frac {\p F_1}{\p p_1}=-8p_2,\quad
q_{2x}=\frac {\p F_1}{\p p_2}=-8p_1+20q_1p_2,\tag 2.10a$$
$$\Psi_{2x}=-\frac {\p F_1}{\p
\Psi_1}=\La^2\Psi_1-q_1\La\Psi_1-q_2\Psi_1,
\tag 2.10b$$
$$p_{1x}=-\frac {\p F_1}{\p q_1}=\frac {35}{128}q_1^4+\frac
{15}{16}q_1^2q_2
-10p_2^2
+\frac {3}{8}q_2^2-\frac 12 <\La\Psi_1, \Psi_1>,\tag 2.10c$$
$$p_{2x}=-\frac {\p F_1}{\p q_2}=\frac {5}{16}q_1^3+\frac {3}{4}q_1q_2
-\frac 12 <\Psi_1, \Psi_1>,\tag 2.10d$$
or equivalently
$$P_x=\theta_0\bigtriangledown F_1,$$
where
$$P=(\Psi_1^T, q_1, q_2, \Psi_2^T, p_1, p_2)^T,\qquad \theta_0=
\left( \matrix 0&I_{(N+2)\times(N+2)}\\-I_{(N+2)\times(N+2)}&0
\endmatrix\right),$$
$$F_1=\frac 12 <\Psi_2, \Psi_2>-\frac 12 <\La^2\Psi_1, \Psi_1>
+\frac 12 q_1<\La\Psi_1, \Psi_1>+\frac 12 q_2<\Psi_1, \Psi_1>$$
$$-8p_1p_2
+10q_1p_2^2-\frac {5}{16}q_1^3q_2-\frac {3}{8}q_1q_2^2
-\frac {7}{128}q_1^5.\tag 2.11$$
The Lax representation for FDHS (2.10) can be deduced from the adjoint
representation (2.2) by using the method in [16,17] which is
sketched as follows. Due to (2.4a), (2.6) and (2.7b), we may define
$$\widetilde b_{m}=\frac 12<\La^{m-5}\Psi_1,\Psi_1>,
\qquad m=5,6,...,$$
which together with (2.4b) and (2.10) yields
$$\widetilde a_{m}=-\frac 12<\La^{m-5}\Psi_1,\Psi_2>,
\quad \widetilde c_{m}=-\frac 12<\La^{m-5}\Psi_2,\Psi_2>, \qquad
m=5,6,....$$
Set
$$\widetilde a_{m}=a_m, \quad \widetilde b_{m}=b_m,\quad
\widetilde c_{m}=c_m, \qquad m=0,1,2,3,4.$$
Then the construction of $\widetilde a_{m}, \widetilde b_{m},
\widetilde c_{m}$
ensures that under (2.10)
$$\widetilde V=\sum_{i=0}^{\infty}\widetilde V_i\la^{-i}, \qquad
\widetilde V_i
=\left( \matrix \widetilde a_i&\widetilde b_i\\\widetilde c_i&-
\widetilde a_i\endmatrix\right),$$
also satisfies (2.2). Notice that
$$\sum_{m=5}^{\infty}\widetilde a_m\la^{-m+4}
=-\frac 12\sum_{m=0}^{\infty}\sum_{j=1}^{N}(\frac
{\la_j}{\la})^m\psi_{1j}\psi_{2j}
=-\frac 12\sum_{j=1}^{N}\frac {\psi_{1j}\psi_{2j}}{\la-\la_j}, $$
set
$$Q\equiv\left( \matrix A(\la)&B(\la)\\C(\la)&-A(\la)\endmatrix\right)
=\la^4\widetilde V,\tag 2.12a$$
we have
$$A(\la)=2p_2\la+2p_1-2q_1p_2-\frac
{1}{2}\sum_{j=1}^{N}\frac{\psi_{1j}\psi_{2j}}{\la-\la_{j}}, \tag 2.12b$$

$$B(\la)=\la^2+\frac 12q_1\la+\frac 38 q_1^2+\frac 12q_2+\frac
{1}{2}\sum_{j=1}^{N}\frac{\psi_{1j}^2}{\la-\la_{j}}, \tag 2.12c$$
$$C(\la)=\la^4-\frac 12q_1\la^3-(\frac 12 q_2+\frac 18q_1^2)\la^2
+(\frac 14q_1^3+\frac 12 q_1q_2-\frac 12<\Psi_1,\Psi_1>)\la$$
$$+\frac 14q_2^2-\frac 5{64}q_1^4-4p_2^2-\frac 12<\La\Psi_1,\Psi_1>
+\frac 12q_1<\Psi_1, \Psi_1>-\frac
{1}{2}\sum_{j=1}^{N}\frac{\psi_{2j}^2}{\la-\la_{j}}. \tag 2.12d$$
Since $\widetilde V$ under (2.10) satisfies (2.2), then $Q$
under (2.10)
satisfies (2.2), too, namely
$$Q_x=[U, Q],\tag 2.13$$
which presents the Lax representation for (2.10). This can also be
verified by a direct
calculation.
The equation (2.13) implies that $\frac 12Tr
Q^2(\la)=A^2(\la)+B(\la)C(\la)$
is the generating function of the integrals of motion for (2.10). We
have
$$A^2(\la)+B(\la)C(\la)=\la^6-F_1\la+F_2+
\sum_{i=1}^{N}\frac{F^{(i)}}{\la-\la_{i}}, \tag 2.14$$
$$F_2=-\frac 12 <\La\Psi_2, \Psi_2>+\frac 12 <\La^3\Psi_1, \Psi_1>
-\frac 14 q_1<\La^2\Psi_1, \Psi_1>$$
$$+(\frac 18 q_1^3+\frac 14q_1q_2-\frac 14<\Psi_1, \Psi_1>)
<\Psi_1, \Psi_1>-\frac 14q_1<\Psi_2, \Psi_2>$$
$$+(\frac 38q_1^2+\frac 12q_2)(-4p_2^2-\frac {5}{64}q_1^4+\frac
{1}{4}q_2^2
-\frac 12<\La\Psi_1, \Psi_1>
+\frac 12 q_1<\Psi_1, \Psi_1>)$$
$$-(\frac 14 q_2
+\frac 1{16} q_1^2)<\La\Psi_1, \Psi_1>
-2p_2<\Psi_1, \Psi_2>+4(p_1-q_1p_2)^2,\tag 2.15$$
$$F^{(i)}=(-2p_2\la_i+2q_1p_2-2p_1)\psi_{1i}\psi_{2i}
-\frac 12(\la_i^2+\frac 12q_1\la_i+\frac 38 q_1^2 +\frac
12q_2)\psi_{2i}^2$$
$$+\frac 12[\la_i^4-\frac 12q_1\la_i^3-(\frac 18 q_1^2 +\frac
12q_2)\la_i^2
+(\frac 14q_1^3+\frac 12q_1q_2-\frac 12<\Psi_1, \Psi_1>)\la_i
+\frac 14q_2^2-4p_2^2$$ $$-\frac 5{64}q_1^4
-\frac 12<\La\Psi_1,\Psi_1>+\frac 12q_1<\Psi_1,\Psi_1>]\psi_{1i}^2
+\frac{1}{4}\sum_{k\neq
i}\frac{(\psi_{1i}\psi_{2k}-\psi_{1k}\psi_{2i})^2}
{\la_k-\la_{i}}, \quad i=1,...,N,\tag 2.16$$
where $F^{(i)}, i=1,...,N, F_1, F_2$ are $N+2$ independent integrals
of motion for
(2.10). By means of the $r$-matrix, it can be shown  that the equation
(2.10) is a finite-dimensional integrable Hamiltonian system
(FDIHS).\par
In order to find the QBH structure for (2.10), we need to use the
modified
system of (2.10). Let us consider the modified Jaulent--Miodek (MJM)
spectral problem [31]
$$\phi_x=U(v,\la)\phi,\quad
 U(v, \lambda)
=\left( \matrix v_0&\la\\\la-v_1&-v_0\endmatrix\right),\quad
\phi=\binom {\phi_{1}}{\phi_{2}},\quad v=\binom{v_0}{v_1}.\tag 2.17$$
The equations (2.2) and (2.3) yield
$$a_{0}=0,\quad b_{0}=1,\quad
b_{1}=\frac{1}{2}v_{1},\quad
a_{1}=v_{0},\quad c_0=1,\quad c_1=-\frac 12 v_1,...$$
$$\left( \matrix 2a_{k+1}\\-b_{k+1}\endmatrix\right)
=L\left( \matrix 2a_k\\-b_{k}\endmatrix\right),
\qquad k=1,2,...,\tag 2.18$$
$$L=\left( \matrix 0&-2v_0+D\\
\frac{1}{4}D+\frac 12D^{-1}v_{0}D&\frac 12v_1+\frac 12D^{-1}v_{1}D
\endmatrix\right).$$
The MJM hierarchy associated with (2.17) can also be written as a
infinite-dimensional
Hamiltonian system
$$v_{t_n}=\left( \matrix v_{0}\\v_{1}\endmatrix\right)_{t_n}
=J\left( \matrix 2a_n\\-b_{n}\endmatrix\right)
=J\frac {\delta H_{n}}{\delta v},\qquad n=1,2,\hdots, \tag 2.19$$
where the Hamiltonian $H_n$ and the Hamiltonian operator $J$ are given
by
$$J=\left( \matrix
\frac 12D&0\\0&-2D\endmatrix\right),\qquad
H_{n}=\frac{-1}{n}[a_{n,x}
-v_1b_n+2b_{n+1}].$$
Also we have
$$\frac {\delta \lambda}{\delta v}=
\binom {2\phi_1\phi_2}{\phi_{1}^{2}}.\tag 2.20$$\par
In the similar way as for (2.7), the constrained flow of (2.19) is
defined by
$$ \Phi_{1,x}=v_0\Phi_{1}+\La\Phi_{2},\quad
\Phi_{2,x}=(\La-v_1)\Phi_{1}-v_0\Phi_2,\tag 2.21a$$
$$\frac {\delta H_l}{\delta v}+
\frac 12\left( \matrix 2<\Phi_1,\Phi_2>\\
<\Phi_1,\Phi_1>\endmatrix\right)=0,\tag 2.21b$$
where $\Phi_i=(\phi_{i1},...,\phi_{iN})^T, i=1,2.$\par
For $l=3$,
$$H_3=-(\frac {1}{4}v_{0x}^2-\frac {1}{16}v_{1x}^2+\frac {1}{4}v_{0}^4
+\frac {5}{64}v_{1}^4-\frac {3}{8}v_{0x}v_1^2-\frac
{3}{8}v_{0}^2v_1^2).$$
By introducing the Jacobi-Ostrogradsky coordinates
$$\widetilde {q_1}=v_1,\qquad \widetilde {q_2}=v_0,\tag 2.22a$$
$$\widetilde {p_1}=-\frac {\delta H_3}{\delta v_{1x}}=-\frac 18 v_{1x},
\quad
\widetilde {p_2}=-\frac {\delta H_3}{\delta v_{0x}}=\frac 12v_{0x}
-\frac 38v_1^2,\tag 2.22b$$
the equations (2.21) for $l=3$ are transformed into a FDHS
$$ \Phi_{1,x}=\frac {\p \widetilde F_1}{\p \Phi_2}
=\widetilde q_2\Phi_1+\La\Phi_2,\qquad
\widetilde q_{1x}=\frac {\p \widetilde F_1}{\p \widetilde p_1}=
-8\widetilde p_1, \quad\widetilde q_{2x}
=\frac {\p \widetilde F_1}{\p \widetilde p_2}=2\widetilde p_2
+\frac 34\widetilde q_1^2, \tag 2.23a$$
$$\Phi_{2,x}=-\frac {\p \widetilde F_1}{\p \Phi_1}=\La\Phi_1
-\widetilde q_1\Phi_1-\widetilde q_2\Phi_2,\tag 2.23b$$
$$\widetilde p_{1x}=-\frac {\p \widetilde F_1}{\p \widetilde q_1}=
-\frac 32\widetilde q_1\widetilde p_2-\frac 34\widetilde q_1\widetilde
q_2^2
-\frac 14\widetilde q_1^3-\frac 12<\Phi_1, \Phi_1>, \tag 2.23c$$
$$\widetilde p_{2x}=-\frac {\p \widetilde F_1}{\p \widetilde q_2}
=\widetilde q_2^3
-\frac 34\widetilde q_1^2\widetilde q_2-<\Phi_1, \Phi_2>, \tag 2.23d$$
or
$$\widetilde P_{x}=\theta_0\bigtriangledown \widetilde F_{1},$$
where
$$\widetilde P=(\Phi_1^T, \widetilde q_{1}, \widetilde q_{2},
\Phi_2^T, \widetilde p_{1}, \widetilde p_{2})^T, $$
$$\widetilde F_{1}=-4\widetilde p_1^2+\widetilde p_2^2
+\frac 34\widetilde q_1^2\widetilde p_2
+\frac 38\widetilde q_1^2\widetilde q_2^2+\frac 1{16}\widetilde q_1^4
-\frac 14\widetilde q_2^4$$
$$+\widetilde q_2<\Phi_1, \Phi_2>+\frac 12<\La\Phi_2, \Phi_2>
-\frac 12<\La\Phi_1, \Phi_1>+
\frac 12\widetilde q_1<\Phi_1, \Phi_1>. $$\par

\subhead {B. The QBH structure for the FDIHS (2.10)}\endsubhead\par
\ \par
We now establish a map relating FDIHS (2.10) to (2.23), then use the
map to construct the second compatible Poisson tensor for the FDIHS
(2.10).\par
It is known [31] that a gauge transformation between the JM and MJM
spectral problem is as follows
$$\psi_1=\phi_1, \quad \psi_2=\la\phi_2+v_0\phi_1, \quad u_1=v_1,
\quad u_0=-v_{0x}-v_0^2, \tag 2.24$$
which, together with (2.9) and (2.22), gives rise to the map relating
(2.10) to (2.23), i.e. $P=M(\widetilde P)$:
$$\Psi_1=\Phi_1, \quad \Psi_2=\La\Phi_2+\widetilde q_2\Phi_1,
\quad q_1=\widetilde q_1,$$
$$q_2=-2\widetilde p_2-\frac 34\widetilde q_1^2-\widetilde q_2^2, \quad
p_1=\widetilde q_1\widetilde p_1+\frac 14\widetilde q_2^3
+\frac 12\widetilde q_2\widetilde p_2-\frac 14<\Phi_1, \Phi_2>,\quad
p_2=\widetilde p_1. \tag 2.25$$
The map $M$ given by (2.25) transforms  all equations in (2.10) except
for (2.10c) into the corresponding equations in (2.23) except for
(2.23c).
In fact, the equation (2.10c) with an additive constant term $c=-
\frac 12\widetilde F_1$ is transformed into (2.23c) under the map
(2.25).
However, the second Poisson tensor constructed later by using the map
(2.25) is valid for an arbitrary $c$, therefore  we can take $c=0$.
The Jacobi $M'$ of the map $M$ take the form
$$M'(\widetilde P)=\left( \matrix I&0&0&0&0&0\\0&1&0&0&0&0
\\0&-\frac 32\widetilde q_1&-2\widetilde q_2&0&0&-2
\\\widetilde q_2I&0&\Phi_1&\La&0&0
\\-\frac 14\Phi_2^T&\widetilde p_1&\frac 34\widetilde q_2^2
+\frac 12\widetilde p_2&
-\frac 14\Phi_1^T&\widetilde q_1&\frac 12\widetilde q_2
\\0&0&0&0&1&0\endmatrix\right).\tag 2.26$$
According to the standard procedure [32], the image of the Poisson
tensor $\theta_0$ for the FDIHS (2.23) under the map $M$ gives rise
to the second compatible Poisson tensor for the FDIHS (2.10). That is
$$\theta_1=M'\theta_0{M'}^T\mid_{P=M(\widetilde P)}=
\left( \matrix 0_{(N+2)\times (N+2)}&A_1\\-A_1^T&B_1\endmatrix\right),
\tag 2.27a$$
$$A_1=\left( \matrix \La&-\frac 14\Psi_1&0_{N\times 1}\\0_{1\times N}
&q_1&1\\2\Psi_1^T&-\frac 12q_2-\frac {15}8 q_1^2&-\frac 32q_1\endmatrix
\right),\quad
B_1=\left( \matrix 0_{N\times N}&\frac 14\Psi_2&0_{N\times 1}\\
-\frac 14\Psi_2^T&0&p_2\\0_{1\times N}&-p_2&0\endmatrix\right).\tag
2.27b$$
Furthermore, by a straightforward calculation, we can show the
following proposition.
\proclaim {Proposition 1}  The system (2.10) possesses the QBH
representation
$$P_x=\theta_0\bigtriangledown F_1=\frac 1{\rho}\theta_1
\bigtriangledown E_1 \tag 2.28$$
where
$$\rho=B(\la)|_{\la=0}=\frac 38q_1^2+\frac 12q_2
-\frac 12<\La^{-1}\Psi_1, \Psi_1>, \tag 2.29a$$
$$E_1=[A^2(\la)+B(\la)C(\la)]|_{\la=0}=F_2-\sum_{i=1}^N\la_i^{-1}F^{(i)}$$

$$=(\frac 3{16}q_1^2+\frac 14q_2)(<\La^{-1}\Psi_2, \Psi_2>
-<\La\Psi_1, \Psi_1>)$$
$$(\frac 3{16}q_1^3+\frac 14q_1q_2)<\Psi_1, \Psi_1>
+2(p_1-q_1p_2)<\La^{-1}\Psi_1, \Psi_2>$$
$$+(2p_2^2+\frac 5{128}q_1^4-\frac 18q_2^2+\frac 14<\La\Psi_1, \Psi_1>
-\frac 14q_1<\Psi_1, \Psi_1>)<\La^{-1}\Psi_1, \Psi_1>$$
 $$+(\frac 38q_1^2+\frac 12q_2)
(\frac 14q_2^2-4p_2^2-\frac 5{64}q_1^4)+4(p_1-q_1p_2)^2$$ $$
+\frac 14[<\La^{-1}\Psi_1, \Psi_2>^2
-<\La^{-1}\Psi_1, \Psi_1><\La^{-1}\Psi_2, \Psi_2>].
\tag 2.29b$$\endproclaim \par

\subhead {C. The Nijenhuis coordinates}\endsubhead\par
\ \par
We now prove that the Nijenhuis coordinates for QBH system (2.28)
are the same as the separated variables defined by means of the Lax
matrix (2.12b). As $\theta_0$ and $\theta_1$ are compatible and
invertible,
the matrix $\theta_1\theta_0^{-1}$ is maximal, it has $N+2$ distinct
eigenvalues $\pmb {\mu}=(\mu_1,...,\mu_{N+2})$. The explicit form of
the canonical transformation from $P$ to the Nijenhuis coordinates
($\pmb {\mu}$, $\pmb {\nu}$) is given in what follows.
The eigenvalues $\mu_1,...,\mu_{N+2}$ are defined by the roots of
the equation
$$f(\la)=\mid \la I-A_1\mid=0, \tag 2.30$$
which, since $A_1$ depends only on $(\Psi_1, q_1, q_2)$, gives rise to
$$\mu_j=f_j(\Psi_1, q_1, q_2), \qquad j=1,...,N+2,\tag 2.31$$
$$\psi_{1j}=g_j(\pmb{\mu}),\quad j=1,...,N, \quad
q_1=g_{N+1}(\pmb{\mu}),
\quad q_2=g_{N+2}(\pmb{\mu}).\tag 2.32$$
Then we introduce the generating function by
$$S=\sum_{j=1}^{N}\psi_{2j}g_j(\pmb{\mu})+p_1g_{N+1}(\pmb{\mu})
+p_2g_{N+2}(\pmb{\mu}),\tag 2.33a $$
such that
$$\psi_{1j}=\frac {\p S}{\p\psi_{2j}}, \quad j=1,...,N,\quad q_1
=\frac {\p S}{\p p_1},\quad q_2=\frac {\p S}{\p p_{2}},\tag 2.33b$$
$$\nu_j=\frac {\p S}{\p\mu_{j}}=\sum_{j=1}^{N}\psi_{2j}\frac {\p g_j}
{\p\mu_j}+p_1\frac {\p  g_{N+1}}{\p\mu_j}+p_2\frac {\p
g_{N+2}}{\p\mu_j},
\quad j=1,...,N+2.\tag 2.33c$$
The equations (2.33b) reconstruct (2.31) or (2.32), the equations
(2.33c)
give the expression for $\nu_j$.
The system (2.10) written in terms of ($\pmb {\mu}$,$\pmb {\nu}$)
can be
shown to be  separable.\par
On the other hand, the separated variables $(\bar{\pmb {\mu}},
\bar{\pmb {\nu}})$ for (2.10)
can be constructed by means of the Lax matrix in the following
way [27,28]. The coordinates $\bar \mu_1,...,\bar\mu_{N+2}$ are
introduced by the zeros of $B(\la)$:
$$B(\la)=\la^2+\frac 12q_1\la+\frac 38 q_1^2+\frac 12q_2+\frac
{1}{2}\sum_{j=1}^{N}\frac{\psi_{1j}^2}{\la-\la_{j}}=\frac
{R(\la)}{K(\la)},
\tag 2.34$$
with
$$R(\la)=\prod_{k=1}^{N+2}(\la-\bar\mu_{k})=
\sum_{k=0}^{N+2}\beta_k\la^{N+2-k}, \quad
K(\la)=\prod_{k=1}^{N}(\la-\la_{k})=
\sum_{k=0}^{N}\alpha_k\la^{N-k}, \tag 2.35$$
$$\alpha_0=1, \quad \alpha_1=-\sum_{j=1}^{N}\la_j,\quad ...,\quad
\alpha_N=(-1)^N\prod_{j=1}^{N}\la_{j},$$
$$\beta_0=1,\quad \beta_1=-\sum_{j=1}^{N+2}\bar{\mu}_j,\quad...,\quad
\beta_{N+2}=(-1)^N\prod_{j=1}^{N+2}\bar{\mu}_{j},$$
and the canonically conjugate coordinates $\bar{\nu_1},...,
\bar{\nu}_{N+2}$ are defined by
$$\bar\nu_k=-A(\bar\mu_k)
=-2p_2\bar\mu_k-2p_1+2q_1p_2+\frac
{1}{2}\sum_{j=1}^{N}\frac{\psi_{1j}\psi_{2j}}{\bar\mu_k-\la_{j}},
\quad k=1,...,N+2. \tag 2.36$$
The FDIHS (2.10) in terms of the coordinates $(\bar{\pmb{\mu}},
\bar{\pmb{\nu}})$ will be  shown to be separable later.
We have the following proposition.
\proclaim {Proposition 2} The Nijenhuis coordinates $(\pmb{\mu},
\pmb{\nu})$ defined by (2.30) and (2.33) are exactly the same as
the separated variables $(\bar{\pmb{\mu}},\bar{\pmb{\nu}})$ defined
by (2.34) and (2.36). The QBH vector field (2.28) is Pfaffian in the
Nijenhuis coordinates.\endproclaim \par
Proof. We first show that
$$f(\la)=B(\la)K(\la)=R(\la). \tag 2.37$$
We denote $f(\la)$ by $f_N(\la;\la_1,...,\la_N)$ in order to prove
(2.37) by induction. Obviously, (2.37) holds for $N=1$. Then we have
by induction
$$f_N(\la;\la_1,...,\la_N)
=\left | \matrix \la-\la_{1}&0&\hdots&0&\frac 14\psi_{11}&0\\
0&\la-\la_{2}&\hdots&0&\frac 14\psi_{12}&0\\
\vdots&\vdots&\ddots&\vdots&\vdots&\vdots\\
0&0&\hdots&\la-\la_{N}&\frac 14\psi_{1N}&0\\
0&0&\hdots&0&\la-q_{1}&-1\\
-2\psi_{11}&-2\psi_{12}&\hdots&-2\psi_{1N}&\frac 12 q_2
+\frac {15}8q_1^2&\la+\frac 32q_1
\endmatrix\right |$$
$$=(\la-\la_1)f_{N-1}(\la;\la_2,...,\la_N)
+\frac 12\psi_{11}^2\prod_{k=2}^{N}(\la-\la_{k})\tag 2.38$$
$$=(\la^2+\frac 12q_1\la+\frac 38 q_1^2+\frac 12q_2+\frac
{1}{2}\sum_{j=2}^{N}\frac{\psi_{1j}^2}{\la-\la_{j}})K(\la)
+\frac {1}{2}\frac{\psi_{11}^2}{\la-\la_{1}}K(\la)=B(\la)K(\la).$$
The equation (2.38) implies that $\la_1,$ similarly $\la_k, k=2,...,N$,
is not the zero of $f(\la)$. Thus (2.37) indicates that $f(\la)$ and
$B(\la)$ have the same zeros, i.e. $\mu_k=\bar\mu_k$.\par
It follows from (2.34) that
$$ \psi_{1j}^2=2\frac{R(\la_j)}{K'(\la_{j})},\qquad
q_1=2(\beta_1-\alpha_1), \tag 2.39a$$
$$\frac 12 q_2+\frac 38q_1^2=\frac 12<\La^{-1}\Psi_{1}, \Psi_1>
+\frac {\beta_{N+2}}{\alpha_N},
\tag 2.39b$$
where the prime denotes differentiation with respect to $\la$.
The equations (2.39a) and (2.39b) yield
$$q_2=2\sum_{j=1}^N\frac{R(\la_j)}{\la_jK'(\la_{j})}-
3(\beta_1-\alpha_1)^2
+2\frac {\beta_{N+2}}{\alpha_N}.\tag 2.39c$$
According to (2.33a), one gets
$$S=\sum_{j=1}^N\psi_{2j}\sqrt{\frac{2R(\la_j)}{K'(\la_{j})}}
+2p_1(\beta_1-\alpha_1)
+p_2[\sum_{j=1}^N\frac{2R(\la_j)}{\la_jK'(\la_{j})}-
3(\beta_1-\alpha_1)^2
+2\frac {\beta_{N+2}}{\alpha_N}].$$
Notice that
$$\frac {\p}{\p \mu_k}\sum_{j=1}^N\psi_{2j}\sqrt{\frac{2R(\la_j)}
{K'(\la_{j})}}
=\sum_{j=1}^N\frac {\psi_{2j}R(\la_j)}{\sqrt{2R(\la_j)K'(\la_{j})}
(\mu_k-\la_j)}
=\frac {1}{2}\sum_{j=1}^{N}\frac{\psi_{1j}\psi_{2j}}{\mu_k-\la_{j}},$$
$$\frac {\p}{\p \mu_k}\sum_{j=1}^N\frac{2R(\la_j)}{\la_jK'(\la_{j})}
=\sum_{j=1}^N\frac{2R(\la_j)}{\la_jK'(\la_{j})(\mu_k-\la_j)}$$
$$=\frac {1}{\mu_k}\sum_{j=1}^N[\frac{2R(\la_j)}{\la_jK'(\la_{j})}
+\frac{2R(\la_j)}{(\mu_k-\la_j)K'(\la_{j})}]
=\frac 1{\mu_k}(<\La^{-1}\Psi_1, \Psi_1>+\sum_{j=1}^{N}
\frac{\psi_{1j}^2}{\mu_k-\la_{j}}),$$
$$\frac {\p \beta_{N+2}}{\p \mu_k}=\frac {\beta_{N+2}}{\mu_k},\qquad
\frac {\p (\beta_1-\alpha_1)^2}{\p \mu_k}=-q_1,\quad
\frac {\p (\beta_1-\alpha_1)}{\p \mu_k}=-1,$$
and using $B(\mu_k)=0$, one finds from (2.33c) that $\nu_k, \mu_k$
satisfy (2.36). Finally, it follows from (2.39b) that
$$\rho=B(\la)|_{\la=0}=\frac 12 q_2+\frac 38q_1^2-\frac
12<\La^{-1}\Psi_{1},
 \Psi_1>=\frac {\beta_{N+2}}{\alpha_N}=\frac {(-1)^N}{\alpha_N}
 \prod_{j=1}^{N+2}\mu_{j}.\tag 2.40$$
This completes the proof.\par
\subhead{D. Comparison of the two methods for
separability}\endsubhead\par
\ \par
For the FDIHS with QBH structure, the separated variables, i.e.
the Nijenhuis coordinates,  can be introduced by the Nijenhuis tensor.
Then the separability of the Hamilton-Jacobi equation for the system
can be shown by varifying the Levi-Civita conditions. For the FDIHS
with some kind of Lax representation, the separated variables can be
found and the separability of the Hamilton Jacobi equation for the
system can be shown by means of the Lax representation.
So far there is not an effective way to define separated variables
for the FDIHSs with some kind of Lax matrices, such as the $3\times 3$
Lax matrices [22] or the Lax matrices admitting dynamical $r$-matrix.
However, if the separated variables can be introduced by the
Lax matrix,
one can further establish the Jacobi inversion problem for the system
by means of the Lax representation. By using the standard Jacobi
inversion technique, the solution to the system can be found.\par
We now use the Lax representation (2.12) to construct the Jacobi
inversion problem for (2.10). Set
$$A^2(\la)+B(\la)C(\la)=\frac {W(\la)}{K(\la)},\qquad
 W(\la)=\sum_{i=0}^{N+6}P_i\la^i, \tag 2.41$$
then $P_i$ are also the integrals of motion for (2.10). By substituting
(2.13) and using (2.14), (2.41) leads to
$$P_{N+6}=1,\quad P_{N+6-i}=\alpha_i,\quad i=1,2,3,4,$$
$$F_1=-P_{N+1}+\alpha_5,\qquad F_2=P_{N}-\alpha_1P_{N+1}
+\alpha_1\alpha_5-\alpha_6,....\tag 2.42$$
The equations (2.34), (2.36) and (2.41) give rise to
$$\nu_k=\sqrt {\frac {W(\mu_k)}{K(\mu_k)}},\qquad k=1,...,N+2, \tag
2.43$$
which indicates that the Hamilton-Jacobi equation is separable.
Replacing $\nu_k$ by $\frac {\p S_k}{\p\mu_k}$ and interpreting
the $P_i$ as integration constants, one gets the generating function
$S$ of the canonical transformation from (2.43)
$$S(\mu_1,...,\mu_{N+2};P_0,...,P_{N+1})=\sum_{k=1}^{N+2}\int^{\mu_k}
\sqrt {\frac {W(\la)}{K(\la)}}d\la. \tag 2.44$$\par
The linearizing coordinates are then
$$Q_i=\frac {\p S}{\p P_i}=\frac 12\sum_{k=1}^{N+2}\int^{\mu_k}
\frac {\la^i}{\sqrt { W(\la)K(\la)}}d\la, \quad i=0,1,...,N+1. \tag
2.45$$
The linear flow induced by (2.10) is then given by (using (2.42))
$$Q_i=\gamma_i+x\frac {\p F_1}{\p P_i}=\gamma_i-x\delta_{i,N+1},
\quad i=0,1,...,N+1, \tag 2.46$$
where $\gamma_i$ are arbitrary constants. Combining the equation
(2.45) with the equation (2.46) leads to the Jacobi inversion problem
for the FDIHS (2.10)
$$\frac 12\sum_{k=1}^{N+2}\int^{\mu_k}\frac {\la^i}{\sqrt {
W(\la)K(\la)}}d
\la
=\gamma_i-x\delta_{i,N+1},\quad i=0,1,...,N+1. \tag 2.47$$
Since $\psi_{1j}, q_1, q_2$ defined by (2.39) are the symmetric
functions
of $\mu_k, k=1,...,N+2$  by using the standard Jacobi inversion
technique
[33], they can be solved in terms of Riemann theta functions from
(2.47).
After having $\psi_{1j}, q_1, q_2$, the $\psi_{2j}, p_1, p_2$ can be
found
by using (2.10a). In this way the solution to (2.10) is obtained.
 \par

\subhead {III. The first family of QBH systems}\endsubhead\par
\ \par
In the following  sections, by using the method described in the
previous
section, we will present QBH representation for some families of FDIHSs
given in [24], and prove the equivalence of two sets of separated
variables.
\par
\subhead {A. The first family of FDIHSs}\endsubhead\par
\ \par
 We first recall the constrained flows of the hierarchy of nonlinear
 evolution
equations (NLEE) associated with  the following polynomial second order
spectral  problem [31]
$$\psi_x=U(u,\la)\psi,\quad
 U(u, \lambda)
=\left( \matrix 0&1\\-\sum_{i=0}^m u_i
\lambda^{i}&0\endmatrix\right),\quad \quad
\psi=\binom {\psi_{1}}{\psi_{2}},\tag 3.1$$
where $u_m=-1, u=(u_{m-1},...,u_0)^T.$
The adjoint representation (2.2) of (3.1) yields
$$a_{0}=\hdots=a_{m}=b_{0}=\hdots=b_{m-1}=0,\quad b_{m}=1,\quad
b_{m+1}=\frac{1}{2}u_{m-1},$$
$$a_{m+1}=-\frac{1}{4}u_{m-1,x},
\quad c_{0}=1,\quad c_{1}=-\frac{1}{2}u_{m-1},\hdots,$$
and in general
$$\left( \matrix b_{k+m}\\\vdots\\b_{k+1}\endmatrix\right)
=L\left( \matrix b_{k+m-1}\\\vdots\\b_{k}\endmatrix\right)
,\tag 3.2a$$
$$ a_k=-\frac 12b_{k,x},\qquad
c_{k}=-\frac{1}{2}b_{k,xx}-\sum_{i=0}^mu_{i}b_{k+i},
\quad k=1,2,\cdots,\tag 3.2b$$
where
$$L=\left( \matrix L_{m-1}&L_{m-2}&\hdots&L_1&L_0\\1&0&\hdots&0&0\\
0&1&\hdots&0&0\\\vdots&\vdots&\ddots&\vdots&\vdots\\0&0&\hdots&1&0
\endmatrix\right),$$
$$L_0=\frac{1}{4}D^2+u_0-\frac 12D^{-1}u_{0,x},\qquad
L_i=u_i-\frac 12D^{-1}u_{i,x},\qquad i=1,\hdots,m-1.$$
The hierarchy of NLEEs associated with (3.1) can be written as an
infinite-dimensional
Hamiltonian system
$$u_{t_n}=\left( \matrix u_{m-1}\\\vdots\\u_0\endmatrix\right)_{t_n}
=J\left( \matrix b_{n+m}\\\vdots\\b_{n+1}\endmatrix\right)
=J\frac {\delta H_{n}}{\delta u},\qquad n=1,2,\hdots, \tag 3.3$$
where the Hamiltonian $H_n$ and the Hamiltonian operator $J$
are defined by
$$J=\left( \matrix
0&0&\hdots&0&2D\\0&0&\hdots&2D&J_{m-1}\\0&0&\hdots&J_{m-1}&J_{m-2}\\
\vdots&\vdots&\ddots&\vdots&\vdots\\2D&J_{m-1}&\hdots&J_1&J_0
\endmatrix\right),$$
$$J_i=-u_{i,x}-2u_iD, \quad i=0,1,\hdots,m-1,\quad
H_{n}=\frac{2}{m-2n-2}\sum_{i=1}^miu_{i}b_{n+i+1}.$$
Under zero boundary condition we have
$$\frac {\delta \lambda}{\delta u}=
(\lambda^{m-1} \psi_{1}^{2},\lambda^{m-2} \psi_{1}^{2},...,
\psi_{1}^{2})^T,\qquad
L\frac {\delta \lambda}{\delta u}=\la\frac {\delta \lambda}{\delta
u}.\tag 3.4$$
\par
Similarly, the constrained flows of the NLEEs (3.3) are defined by [24]
$$ \Psi_{1,x}=\Psi_{2},\qquad
\Psi_{2,x}=\La^m\Psi_{1}-\sum_{i=0}^{m-1}u_i\La^i\Psi_1,\tag 3.5a$$
$$\frac {\delta H_{l}}{\delta u}-\frac 12
\sum_{j=1}^{N}\frac {\delta \lambda_{j}}{\delta u}=
\left( \matrix b_{m+l}\\\vdots\\b_{l+1}\endmatrix\right)
-\frac 12\left( \matrix
<\La^{m-1}\Psi_1,\Psi_1>\\\vdots\\<\Psi_1,\Psi_1>
\endmatrix\right)=0.\tag 3.5b$$\par
For $l=m$, (3.5b) leads to
$$u_{m-k}=\sum_{j=1}^k(-1)^{j-1}\frac
{j+1}{2^j}\sum_{l_1+\hdots+l_j=k-j}
<\La^{l_1}\Psi_1,\Psi_1>\hdots<\La^{l_j}\Psi_1,\Psi_1>,$$
$$ \quad k=1,\hdots,m,\tag 3.6$$
where $l_1\geq 0,...,l_j\geq 0.$ By substituting (3.6) into (3.5a), the
first constrained flow of (3.3) can be written as a canonical FDHS
$$ \Psi_{1,x}=\frac {\p F_0}{\p \Psi_2},\qquad
\Psi_{2,x}=-\frac {\p F_0}{\p \Psi_1},\tag 3.7$$
or
$$P_x=\theta_0\bigtriangledown F_0,$$
where
$$P=(\Psi_1^T, \Psi_2^T)^T,\qquad \theta_0=\left(\matrix 0&
I_{N\times N}\\-I_{N\times N}&0\endmatrix\right),$$
$$F_{0}=\frac 12<\Psi_2,\Psi_2>+\sum_{j=0}^m(\frac
{-1}{2})^{j+1}\sum_{l_1+\hdots+l_{j+1}=m-j}
<\La^{l_1}\Psi_1,\Psi_1>\hdots<\La^{l_{j+1}}\Psi_1,\Psi_1>.$$\par
The entries of the Lax matrix for (3.7) are given by [24]
$$A(\la)=-\frac
{1}{2}\sum_{j=1}^{N}\frac{\psi_{1j}\psi_{2j}}{\la-\la_{j}}, \qquad
B(\la)=1+\frac
{1}{2}\sum_{j=1}^{N}\frac{\psi_{1j}^2}{\la-\la_{j}}, \tag 3.8a$$
$$C(\la)=\la^m+\sum_{k=1}^m\la^{m-k}
\sum_{j=1}^k(-\frac 12)^{j}\sum_{l_1+\hdots+l_j=k-j}
<\La^{l_1}\Psi_1,\Psi_1>\hdots<\La^{l_j}\Psi_1,\Psi_1>$$ $$-\frac
{1}{2}\sum_{j=1}^{N}\frac{\psi_{2j}^2}{\la-\la_{j}}. \tag 3.8b$$
We have
$$A(\la)^2+B(\la)C(\la)=\la^m+
\sum_{i=1}^{N}\frac{F^{(i)}}{\la-\la_{i}}, \tag 3.9$$
$$F^{(i)}=
\frac 12[\la_i^m+\sum_{k=1}^m\la_i^{m-k}
\sum_{j=1}^k(-\frac 12)^{j}
\sum_{l_1+\hdots+l_j=k-j}
<\La^{l_1}\Psi_1,\Psi_1>\hdots<\La^{l_j}\Psi_1,\Psi_1>]\psi_{1i}^2$$
$$-\frac 12\psi_{2i}^2
+\frac{1}{4}\sum_{k\neq
i}\frac{(\psi_{1i}\psi_{2k}-\psi_{1k}\psi_{2i})^2}
{\la_k-\la_{i}}, \quad i=1,...,N,$$
where $F^{(i)}, i=1,...,N,$ are independent integrals of motion for
(3.7) and $F_0=\sum_{i=0}^NF^{(i)}$. It can be shown that the system
(3.7) is integrable in the Liouvill's sense. The systems with
$m=1,2,...$
 give rise to a family of FDIHSs which include the well-known Garnier
 system as the first member (m=1). This family of FDIHSs was first
 given in [34].\par

In order to find the QBH structure for (3.7), we need to
consider the following modified polynomial second order spectral
problem [31]
$$\phi_x=U(v,\la)\phi,\quad
 U(v, \lambda)
=\left( \matrix v_0&\la\\-\sum_{i=1}^{m} v_i
\lambda^{i-1}&-v_0\endmatrix\right),\quad \quad
\phi=\binom {\phi_{1}}{\phi_{2}},\tag 3.10$$
where $v_m=-1, v=(v_{0},...,v_{m-1})^T.$
The equations (2.2) and (2.3) yield
$$a_{0}=\hdots=a_{m-2}=b_{0}=\hdots=b_{m-3}=0,\quad b_{m-2}=1,\quad
b_{m-1}=\frac{1}{2}v_{m-1},$$
$$a_{m-1}=v_{0},
\quad c_{0}=1,\quad c_{1}=-\frac{1}{2}v_{m-1},\hdots,$$
and in general
$$\left( \matrix 2a_{k+1}\\-b_{k+1}\\\vdots\\-b_{k+m-1}
\endmatrix\right)
=L\left( \matrix 2a_k\\-b_{k}\\\vdots\\-b_{k+m-2}\endmatrix\right),
\qquad \tag 3.11a$$
$$c_{k+1}=a_{k,x}-\sum_{i=1}^mv_{i}b_{k+i-1},
\qquad k=1,2,\cdots,\tag 3.11b$$
where
$$L=\left( \matrix 0&-2v_0+D&0&\hdots&0&0\\
0&0&1&\hdots&0&0\\
\vdots&\vdots&\vdots&\ddots&\vdots&\vdots\\0&0&0&\hdots&0&1\\
L_{0}&L_{1}&L_2&\hdots&L_{m-2}&L_{m-1}
\endmatrix\right),$$
$$L_0=\frac{1}{4}D+\frac 12D^{-1}v_{0}D,\qquad
L_i=\frac 12v_i+\frac 12D^{-1}v_{i}D,\qquad i=1,\hdots,m-1.$$
The hierarchy of NLEEs associated with (3.10) is
$$v_{t_n}=\left( \matrix v_{0}\\\vdots\\v_{m-1}\endmatrix\right)_{t_n}
=J\left( \matrix 2a_n\\-b_{n}\\\vdots\\-b_{n+m-2}\endmatrix\right)
=J\frac {\delta H_{n}}{\delta u},\qquad n=1,2,\hdots, \tag 3.12$$
where the Hamiltonian $H_n$ and the Hamiltonian operator $J$ are given
by
$$J=\left( \matrix
\frac 12D&0&0&\hdots&0&0\\0&J_2&J_3&\hdots&J_{m-1}&-2D\\
\vdots&\vdots&\vdots&\ddots&\vdots&\vdots\\0&J_{m-1}&-2D&\hdots&0&0
\\0&-2D&0&\hdots&0&0\endmatrix\right),$$
$$J_i=v_{i,x}+2v_iD, \quad i=0,1,\hdots,m-1,\quad
H_{n}=\frac{2}{m-2n-2}[a_{n,x}
-\sum_{i=1}^miv_{i}b_{n+i-1}].$$
Also we have
$$\frac {\delta \lambda}{\delta u}=
(2\phi_1\phi_2,  \phi_{1}^{2},\la\phi_{1}^{2},...
,\la^{m-2}\phi_{1}^{2})^T
.\tag 3.13$$
\par
The constrained flows of (3.12) are defined by
$$ \Phi_{1,x}=v_0\Phi_{1}+\La\Phi_{2},\quad
\Phi_{2,x}=(\La^{m-1}-\sum_{i=1}^{m-1}v_i\La^{i-1})\Phi_{1}-v_0\Phi_2,
\tag 3.14a$$
$$\frac {\delta H_{l}}{\delta v}+\frac 12
\sum_{j=1}^{N}\frac {\delta \lambda_{j}}{\delta v}=
\left( \matrix
2a_{l}\\-b_{l}\\\vdots\\-b_{l+m-2}\endmatrix\right)
+\frac 12\left( \matrix 2<\Phi_1,\Phi_2>\\
<\Phi_1,\Phi_1>\\
\vdots\\<\La^{m-2}\Phi_1,\Phi_1>
\endmatrix\right)=0.\tag 3.14b$$\par
For $l=m-1$, (3.14b) leads to
$$v_0=-\frac 12<\Phi_1,\Phi_2>,\tag 3.15a$$
$$v_{m-k}=\sum_{j=1}^k(-1)^{j-1}\frac
{j+1}{2^j}\sum_{l_1+\hdots+l_j=k-j}
<\La^{l_1}\Phi_1,\Phi_1>\hdots<\La^{l_j}\Phi_1,\Phi_1>, $$
$$\quad k=1,\hdots,m-1.\tag
3.15b$$
By substituting (3.15) into (3.14a), the
first constrained flow of NLEE (3.12) can be written as a canonical
FDHS
$$ \Phi_{1,x}=\frac {\p \widetilde F_0}{\p \Phi_2},\qquad
\Phi_{2,x}=-\frac {\p \widetilde F_0}{\p \Phi_1},\tag 3.16a$$
or
$$\widetilde P_x=\theta_0\bigtriangledown \widetilde F_0,$$
where
$$\widetilde P=(\Phi_1^T, \Phi_2^T)^T,\qquad \theta_0=\left(\matrix 0
&I_{N\times N}\\-I_{N\times N}&0\endmatrix\right),$$
$$\widetilde F_{0}=\frac 12<\La\Phi_2,\Phi_2>
-\frac 14<\Phi_1,\Phi_2>^2
$$ $$+\sum_{j=1}^m(-\frac 12)^j\sum_{l_1+\hdots+l_{j}=m-j}
<\La^{l_1}\Phi_1,\Phi_1>\hdots<\La^{l_{j}}\Phi_1,\Phi_1>.\tag 3.16b$$
\ \par
\subhead {B.  The QBH structure for the family of FDIHD (3.7)}
\endsubhead\par
\ \par
It is known [31] that the gauge transformation between the spectral
problems (3.1) and (3.10) is given by
$$\psi_1=\phi_1, \quad \psi_2=\la\phi_2+v_0\phi_1, $$
$$\quad u_i=v_i,\quad i=1,...,m-1, \quad u_0=-v_{0x}-v_0^2, \tag 3.17$$
which, together with (3.6) and (3.15), gives rise to the map relating
(3.7) to (3.16), i.e. $P=M(\widetilde P)$:
$$\Psi_1=\Phi_1, \quad \Psi_2=\La\Phi_2-\frac 12<\Phi_1, \Phi_2>\Phi_1.
\tag 3.18$$
In fact the map $M$ transforms the first equation and the second
equation
with an additive term  , $-c\Psi_1 (c=\widetilde F_0)$,  in (3.7) into
the corresponding equations in (3.16). Since the $\theta_1$ constructed
in the following is valid for an arbitrary $c$, so we can take $c=0$.
The Jacobi $M'$ of the map $M$ takes the form
$$M'(\widetilde P)=\left( \matrix I_{N\times N}&0_{N\times N}\\
-\frac 12<\Phi_1, \Phi_2>I_{N\times N}-\frac 12\Phi_1\Phi_2^T&\La
-\frac 12\Phi_1\Phi_1^T
\endmatrix\right).\tag 3.19$$
Then the second compatible Poisson tensor for the vector field (3.7) is
$$\theta_1=M'\theta_0{M'}^T\mid_{P=M(\widetilde P)}=
\left( \matrix 0_{N\times N}&A_1\\-A_1^T&B_1\endmatrix\right),\tag
3.20$$
$$A_1=\La-\frac 12\Psi_1\Psi_1^T,\qquad B_1=\frac 12\Psi_2\Psi_1^T
-\frac 12\Psi_1\Psi_2^T.$$
By a straightforward calculation, we have the following proposition.
\proclaim {Proposition 3} The system (3.7) possesses the QBH
representation
$$P_x=\theta_0\bigtriangledown F_0=\frac 1{\rho}\theta_1
\bigtriangledown E_1 \tag 3.21a$$
where
$$\rho=B(\la)|_{\la=0}=1-\frac 12<\La^{-1}\Psi_1, \Psi_1>, \tag 3.21b$$
$$E_1=[A^2(\la)+B(\la)C(\la)]|_{\la=0}=-\sum_{i=1}^N\la_i^{-1}F^{(i)}$$
$$=
\frac 12<\La^{-1}\Psi_2,\Psi_2>
+\frac 14[<\La^{-1}\Psi_1, \Psi_2>^2-<\La^{-1}\Psi_1, \Psi_1><\La^{-1}
\Psi_2, \Psi_2>]$$
$$+\sum_{j=0}^m(-\frac 12)^{j+1}\sum_{l_1+\hdots+l_{j+1}=m-j}
<\La^{l_1}\Psi_1,\Psi_1>\hdots<\La^{l_{j+1}-1}\Psi_1,\Psi_1>.\tag
3.21c$$
\endproclaim \par

\subhead {C. The Nijenhuis coordinates}\endsubhead\par
\ \par
In the same way as for (2.30)-(2.33),
the eigenvalues of the Nijenhuis tensor $\mu_1,...,\mu_{N}$ are defined
by the roots of the equation
$$f(\la)=\mid \la I-A_1\mid=0, \tag 3.22a$$
which gives
$$\psi_{1j}=g_j(\pmb{\mu})\quad j=1,...N.$$
Then one defines
$$\nu_j=\frac {\p S}{\p\mu_{j}}=\sum_{j=1}^{N}\psi_{2j}\frac {\p g_j}
{\p\mu_j},\quad j=1,...,N.\tag 3.22b$$\par
On the other hand, the generalized elliptic coordinates
$(\bar{\pmb{\mu}},
\bar{\pmb{\nu}})$
are defined by means of the Lax matrix in the following way [24]. The
coordinates $\bar \mu_1,...,\bar\mu_{N}$ are introduced by the zeros
of $B(\la)$:
$$B(\la)=1+\frac
{1}{2}\sum_{j=1}^{N}\frac{\psi_{1j}^2}{\la-\la_{j}}=\frac {R(\la)}
{K(\la)}, \tag 3.23a$$
where $K(\la)$ is defined by (2.35) and
$$R(\la)=\prod_{k=1}^{N}(\la-\bar\mu_{k})=
\sum_{k=0}^{N}\beta_k\la^{N-k}, \tag 3.23b$$
$$\beta_0=1,\quad \beta_1=-\sum_{j=1}^{N}\bar\mu_j,...,\quad
\beta_{N}=(-1)^N\prod_{j=1}^{N}\bar\mu_{j},$$
and the canonically conjugate coordinates $\bar\nu_1,...,\bar\nu_{N}$
are defined by
$$\bar\nu_k=-A(\bar\mu_k)=
\frac
{1}{2}\sum_{j=1}^{N}\frac{\psi_{1j}\psi_{2j}}{\bar\mu_k-\la_{j}},
\quad k=1,...,N. \tag 3.23c$$\par
We have the following proposition.
\proclaim {Proposition 4} The Nijenhuis coordinates $(\pmb {\mu},
\pmb {\nu})$ defined by (3.22) are exactly the same as the generalized
elliptic coordinates $(\bar{\pmb {\mu}},\bar{\pmb {\nu}})$ defined by
(3.23). The QBH vector field (3.21) is Pfaffian in the Nijenhuis
coordinates.\endproclaim \par
Proof.
Similarly, we have by induction
$$f_N(\la;\la_1,...,\la_N)=|\la I-A_1|$$
$$=\left | \matrix \la-\la_{1}+\frac 12\psi_{11}^2&\frac 12\psi_{11}
\psi_{12}&\hdots&\frac 12\psi_{11}\psi_{1N}\\\frac 12\psi_{12}\psi_{11}
&\la-\la_2+\frac 12\psi_{12}^2&\hdots&\frac 12\psi_{12}\psi_{1N}\\
\vdots&\vdots&\ddots&\vdots\\
\frac 12\psi_{1N}\psi_{11}&\frac 12\psi_{1N}\psi_{12}&\hdots
&\la-\la_{N}
+\frac 12\psi_{1N}^2
\endmatrix\right |$$
$$=(\la-\la_1)f_{N-1}(\la;\la_2,...,\la_N)+\frac 12\psi_{11}^2
\prod_{k=2}^{N}(\la-\la_{k})=B(\la)K(\la),\tag 3.24$$
which shows that $\mu_k=\bar\mu_k$.
It follows from (3.23a) that
$$ \psi_{1j}^2=2\frac{R(\la_j)}{K'(\la_{j})},\qquad j=1,...,N.
\tag 3.25$$
Thus we have
$$\nu_k=\sum_{j=1}^N\psi_{2j}\frac {\p}{\p \mu_k}
\sqrt{\frac{2R(\la_j)}{K'(\la_{j})}}
=\sum_{j=1}^N\frac {\psi_{2j}R(\la_j)}{\sqrt{2R(\la_j)K'(\la_{j})}
(\mu_k-\la_j)}
=\frac {1}{2}\sum_{j=1}^{N}\frac{\psi_{1j}\psi_{2j}}{\mu_k-\la_{j}},
\tag 3.26$$
which implies that $\nu_k=\bar\nu_k$, since $\mu_k=\bar\mu_k$.
Finally, it is found from (3.23a) that
$$\rho=B(\la)|_{\la=0}=1-\frac 12<\La^{-1}\Psi_{1}, \Psi_1>=
\frac {\beta_{N}}{\alpha_N}.$$
This completes the proof.\par
 \ \par
\subhead {IV.  The second family of QBH systems} \endsubhead\par
\ \par
For $l=m+1$, it is found from (3.5b) [24] that
$$u_{m-k}=(-\frac 12)^ku_{m-1}^k$$
$$+\sum_{i=0}^{k-2}u_{m-1}^i\sum_{j=1}^{[\frac {k-i}2]}E_{i,j}
\sum_{l_1+\hdots+l_j=k-i-2j}
<\La^{l_1}\Psi_1,\Psi_1>\hdots<\La^{l_j}\Psi_1,\Psi_1>, \quad k=2,
\hdots,m,\tag
4.1a$$
$$L_0u_{m-1}=
<\La^{m-1}\Psi_1,\Psi_1>-\sum_{i=1}^{m-1}L_i
<\La^{i-1}\Psi_1,\Psi_1>,\tag 4.1b$$
where
$$E_{i,j}=-(i+j+1)\beta_{i,j},\qquad
\beta_{i,j}=(-\frac 12)^{i+j}\frac {(i+j)!}{i!j!}.
 $$\par
Denote
$$q=u_{m-1},\qquad p=-\frac 18u_{m-1,x}. $$
 By substituting (4.1a), (3.5a) and (4.1b) become a canonical FDHS
$$P_x=\theta_0\bigtriangledown F_1,\tag 4.2a$$
where
$$P=(\Psi_1^T, q, \Psi_2^T, p)^T,\qquad \theta_0=\left(\matrix 0
&I_{(N+1)\times (N+1)}\\-I_{(N+1)\times (N+1)}&0\endmatrix\right),$$
$$F_{1}=\frac 12<\Psi_2,\Psi_2>+(-\frac 12q)^{m+2}-4p^2
$$ $$+\sum_{i=0}^mq^i\sum_{j=1}^{[\frac {m+2-i}2]}\beta_{i,j}
\sum_{l_1+\hdots+l_{j}=m+2-i-2j}
<\La^{l_1}\Psi_1,\Psi_1>\hdots<\La^{l_{j}}\Psi_1,\Psi_1>.\tag 4.2b$$
The entries of the Lax matrix $Q$ for (4.2) are of the form [24]
$$A(\la)=2p-\frac
{1}{2}\sum_{j=1}^{N}\frac{\psi_{1j}\psi_{2j}}{\la-\la_{j}}, \qquad
B(\la)=\la+\frac 12q+\frac
{1}{2}\sum_{j=1}^{N}\frac{\psi_{1j}^2}{\la-\la_{j}}, \tag 4.3a$$
$$C(\la)=\sum_{k=0}^{m+1}\la^{m+1-k}\tilde c_k
-\frac {1}{2}\sum_{j=1}^{N}\frac{\psi_{2j}^2}{\la-\la_{j}}, \tag 4.3b$$
where
$$\tilde c_k=(-\frac 12q)^{k}
+\sum_{i=0}^{k-2}q^i \sum_{j=1}^{[\frac {k-i}2]}\beta_{i,j}
\sum_{l_1+\hdots+l_{j}=k-i-2j}
<\La^{l_1}\Psi_1,\Psi_1>\hdots<\La^{l_{j}}\Psi_1,\Psi_1>,$$ $$
\quad k=1,...,m+1, \tag 4.3c$$
$$\tilde c_{m+2+k}=-\frac 12<\La^{k}\Psi_2,\Psi_2>, \quad k=0,1,....
\tag 4.3d$$
Similarly, the equality
$$A^2(\la)+B(\la)C(\la)=\la^{m+2}-F_1+
\sum_{i=1}^{N}\frac{F^{(i)}}{\la-\la_{i}}, \tag 4.4$$
$$F^{(i)}=-2p\psi_{1i}\psi_{2i}-\frac 12(\la_i+\frac 12q)\psi_{2i}^2
+\frac 12\sum_{k=0}^{m+1}\tilde c_k\la_i^{m+1-k}
\psi_{1i}^2$$
$$+\frac{1}{4}\sum_{k\neq
i}\frac{(\psi_{1i}\psi_{2k}-\psi_{1k}\psi_{2i})^2}
{\la_k-\la_{i}}, \quad i=1,...,N,$$
determines $N+1$  independent integrals of
motion  $F_0, F^{(i)}, i=1,...,N,$ for the FDHS (4.2). The systems
(4.2) for $m=1,2,...,$ give the second family of FDIHSs. By taking
$m=1$ (4.2) gives rises to the multidimensional Henon-Heiles system.
 The system (4.2) was also studied by a recurrence relation in [35],
 however no explicit expressions like (4.2b) and (4.3) were given in
 that paper.\par
In the exactly the same way as we did in the previous section, we can
obtain another FDHS from (3.14) for $l=m$, find the map relating this
FDHS to the FDHS (4.2) and finally, by using this map, obtain the
second compatible Poisson tensor for the vector field for (4.2)
$$\theta_1=
\left( \matrix 0_{(N+1)\times (N+1)}&A_1\\-A_1^T&B_1\endmatrix\right),
\tag 4.5$$
$$A_1=\left( \matrix\La&-\frac 14\Psi_1\\2\Psi_1^T&-\frac 12q\endmatrix
\right),\qquad B_1=
\left (\matrix0_{N\times N}&\frac 14\Psi_2\\-\frac
14\Psi_2^T&0\endmatrix
\right).$$
By a straightforward calculation, we can show the following proposition.

\proclaim {Proposition 5} The system (4.2) possesses the QBH
representation
$$P_x=\theta_0\bigtriangledown F_1=\frac 1{\rho}\theta_1
\bigtriangledown
E_1 \tag 4.6$$
where
$$\rho=B(\la)|_{\la=0}=\frac 12q-\frac 12<\La^{-1}\Psi_1, \Psi_1>,
\tag 4.7a$$
$$E_1=[A^2(\la)+B(\la)C(\la)]|_{\la=0}=-F_1-\sum_{i=1}^N
\la_i^{-1}F^{(i)}$$
$$=2p<\La^{-1}\Psi_1,\Psi_2>+\frac 14q<\La^{-1}\Psi_2,\Psi_2>
+4p^2-(-\frac 12q)^{m+2}$$
 $$-\sum_{i=0}^mq^i\sum_{j=1}^{[\frac {m+2-i}2]}\beta_{i,j}
\sum_{l_1+\hdots+l_{j}=m+2-i-2j}
<\La^{l_1}\Psi_1,\Psi_1>\hdots<\La^{l_{j}}\Psi_1,\Psi_1>$$
$$+\frac 14[<\La^{-1}\Psi_1, \Psi_2>^2-<\La^{-1}\Psi_1, \Psi_1>
<\La^{-1}\Psi_2, \Psi_2>]
-\frac 12\sum_{i=0}^{m+1}q^i\sum_{j=0}^{[\frac {m+1-i}2]}\beta_{i,j}$$
$$\times
\sum_{l_1+\hdots+l_{j+1}=m+1-i-2j}
<\La^{l_1}\Psi_1,\Psi_1>\hdots<\La^{l_j}\Psi_1,\Psi_1>
<\La^{l_{j+1}-1}\Psi_1,\Psi_1>
.\tag 4.7b$$\endproclaim \par
In the same way,
$\mu_1,...,\mu_{N+1}$ in the Nijenhuis coordinates are defined by the
roots of the equation
$$f(\la)=\mid \la I-A_1\mid=0, \tag 4.8a$$
which gives
$$\psi_{1j}=g_j(\pmb {\mu})\quad j=1,...N, \quad q=g_{N+1}(\pmb\mu).$$
Then one defines
$$\nu_j=\frac {\p S}{\p\mu_{j}}=\sum_{j=1}^{N}\psi_{2j}\frac {\p g_j}
{\p\mu_j}+p
\frac {\p g_{N+1}}{\p\mu_j},\quad j=1,...,N+1.\tag 4.8b$$\par
On the other hand, the generalized parabolic coordinates
$(\bar{\pmb{\mu}},\bar{\pmb{\nu}})$
are defined by means of the Lax matrix in the following way [24].
The coordinates $\bar \mu_1,...,\bar\mu_{N+1}$ are introduced by
the zeros of $B(\la)$:
$$B(\la)=\la+\frac 12q+\frac
{1}{2}\sum_{j=1}^{N}\frac{\psi_{1j}^2}{\la-\la_{j}}=\frac {R(\la)}
{K(\la)}, \tag 4.9a$$
where $K(\la)$ is defined by (2.35) and $R(\la)$ by
$$R(\la)=\prod_{k=1}^{N+1}(\la-\bar\mu_{k})=
\sum_{k=0}^{N+1}\beta_k\la^{N+1-k}, $$
$$\beta_0=1,\quad \beta_1=-\sum_{j=1}^{N}\bar\mu_j,...,\quad
\beta_{N+1}=(-1)^{N+1}\prod_{j=1}^{N+1}\bar\mu_{j},$$
and the canonically conjugate coordinates $\bar\nu_1,...,\bar\nu_{N+1}$
are defined by
$$\bar\nu_k=-A(\bar\mu_k)=
-2p+\frac
{1}{2}\sum_{j=1}^{N}\frac{\psi_{1j}\psi_{2j}}{\bar\mu_k-\la_{j}},
\quad k=1,...,N+1. \tag 4.9b$$
We have the following proposition.
\proclaim {Proposition 4} The Nijenhuis coordinates $(\pmb{\mu},
\pmb{\nu})$ defined by (4.8) are exactly the same as the generalized
parabolic coordinates $(\bar{\pmb{\mu}},\bar{\pmb{\nu}})$ defined by
(4.9). The QBH vector field (4.6) is Pfaffian in the Nijenhuis
coordinates.\endproclaim \par
Proof. In a similar way, we can show by induction that
$$f(\la)=B(\la)K(\la). \tag 4.10$$
It follows from (4.9a) that
$$ \psi_{1j}^2=2\frac{R(\la_j)}{K'(\la_{j})},\qquad j=1,...,N,$$
$$q=<\La^{-1}\Psi_1, \Psi_1>+2\frac {\beta_{N+1}}{\alpha_N}=
\sum_{j=1}^N\frac {2R(\la_j)}{\la_jK'(\la_{j})}+2\frac {\beta_{N+1}}
{\alpha_N}.
\tag 4.11$$
Then it is similar to find that
$$\nu_k=-A(\mu_k),$$
$$\rho=B(\la)|_{\la=0}=\frac 12q-\frac 12<\La^{-1}\Psi_{1}, \Psi_1>
=\frac {\beta_{N+1}}{\alpha_N}.$$
This completes the proof.\par
\ \par
\subhead {V. Concluding remarks}\endsubhead\par
\ \par
 In the exactly same way as we did in the previous two sections, we
 can construct the third family of QBH systems from the constrained
 flows (3.5) for $l=m+2, m=1,2,....$ The QBH system (2.28) is just
 the second member $(m=2)$ in the third family of QBH systems,  and
 $\theta_1$ and $\rho$ given by (2.27) and (2.29a) are the second
 compatible Poisson tensor and the integrating factor for the third
 family of QBH systems. \par
In general, the constrained flow (3.5) for $l=m+k$ can be transformed
into a FDIHS by introducing the Jacobi-Ostrogradsky coordinates. Under
the map relating this FDIHS to that obtained from the modified
constrained flow (3.14) for $l=m+k-1$, the image of the Poisson
tensor $\theta_0$ for the latter gives rise to the second compatible
Poisson tensor $\theta_1$ for the former. In this way, for each $k$
we can obtain a family of QBH systems with $m=1,2,....$  The results
obtained in the previous sections suggest the following conjecture:
each family of QBH systems ($l=m+k, m=1,2,...$) shares the same
$\theta_1$ and $\rho$ for the QBH structure
$$\theta_0\bigtriangledown F_1=\frac 1{\rho}\theta_1\bigtriangledown
E_1,$$
and, in general, by means of the Lax matrix $Q=\left( \matrix A(\la)
&B(\la)\\C(\la)&-A(\la)\endmatrix\right)$ and the expression
$$A^2(\la)+B(\la)C(\la)=
\sum_{i=0}^{m+2k} \overline F_i\la^{i}+
\sum_{i=1}^{N}\frac{F^{(i)}}{\la-\la_{i}},$$
we have
$$\rho=B(\la)|_{\la=0},\quad E_1=[A^2(\la)+B(\la)C(\la)]|_{\la=0}
=\overline F_0-
\sum_{i=1}^{N}F^{(i)}\la_{i}^{-1}.$$
 For $k=1,2,..,$ we find an infinite number of families of QBH systems.
 Furthermore we can show in a similar way that the Nijenhuis
coordinates
 introduced by the Nijenhuis tensor are exactly the same as the
separated
 variables defined by means  of the Lax matrix for the QBH system
in the
 family, and each QBH vector field is Pfaffian in the Nijenhuis
coordinates.
 \par

\  \par

\subhead {Acknowledgment}\endsubhead\par
This work was supported by the Chinese Basic Research Project
``Nonlinear Science'', the City University of Hong Kong and the
Research Grants Council of Hong Kong.
 One of the authors (Y.B.Zeng) wishes to express his gratitude to
Department of Mathematics of the City University of Hong Kong for warm
hospitality.
\par
\  \par

\subhead {References}\endsubhead \par
\item {1.} F. Magri, J. Math. Phys. 19, 1156 (1978).\par
\item {2.} M. Antonowicz, A. P. Fordy and S. Rauch-Wojciechowski,
Phys. Lett. A 124, 143 (1987).\par
\item {3.} M. Antonowicz and S. Rauch-Wojciechowski, J. Phys. A 24,
5043 (1991).\par
\item {4.} M. Antonowicz and S. Rauch-Wojciechowski, Phys. Lett.
A 163, 167 (1992).\par
\item {5.} M. Antonowicz and S. Rauch-Wojciechowski, J. Math. Phys.
33, 2115 (1992).\par
\item {6.} Yunbo Zeng, J. Phys. A 24, L11 (1993).\par
\item {7.} Yunbo Zeng, J. Math. Phys. 34, 4742 (1993).\par
\item {8.} M. Blaszak, J. Phys. A 26, 5985 (1993).\par
\item {9.} R. Caboz, V. Ravoson and L. Gavrilov, J. Phys. A 24,
L523 (1991).\par
\item {10.} R. Brouzet, R. Caboz, J. Rabenivo and V. Ravoson,
J. Phys. A 29, 2069 (1996).\par
\item {11.} F. Magri and T. Marsico, Electromagnetism and Geometrical
structures, ed G. Ferrarese, Bologna: Pitagora, 1996, p207.\par
\item {12.} C. Morosi and G. Tondo, J. Phys. A 30, 2799 (1997).\par
\item {13.} M. Blaszak, J. Math. Phys. 39, 3213 (1998).\par
\item {14.} J. Rabenivo, J. Phys. A 34, 7113 (1998).\par
\item {15.} Yunbo Zeng, Phys. Lett. A 160, 541 (1991).\par
\item {16.} Yunbo Zeng and Yishen Li, J. Phys. A 26, L273 (1993).\par
\item {17.} Yunbo Zeng, Physica D73, 171 (1994).\par
\item {18.} Yunbo Zeng, J. Phys. A 24, L1065 (1991).\par
\item {19.} W.X. Ma and W. Strampp, Phys. Lett. A 185, 277 (1994).\par
\item {20.} W.X. Ma, J. Phys. Soc. Jpn. 64, 1085 (1995).\par
\item {21.} W.X. Ma, B. Fuchssteiner and W. Oevel, Physica A 233, 331
(1996).\par
\item {22.} W.X. Ma, Q. Ding, W. G. Zhang and B. Q. Lu, IL Nuovo
Cimento B 111, 1135 (1996).\par
\item {23.} O. Ragnisco and S. Rauch-Wojciechowski, Inverse Problems 8,
245 (1992).\par
\item {24.} Yunbo Zeng and Runlian Lin, Families of dynamical
$r$-matrices
and Jacobi inversion
problem for nonlinear evolution equatons, J. Math. Phys., 39, 5964
(1998)\par
\item {25.} E. K. Sklyanin, Prog. Theor. Phys. Suppl. 118, 35
(1995).\par
\item {26.} J. Harnad and P. Winternitz, Commun. Math. Phys. 172, 263
(1995).\par
\item {27.} Yunbo Zeng, J. Phys. A 30, 3719 (1997).\par
\item {28.} Yunbo Zeng, J. Phys. Society of Japan 66, 2277 (1997).\par
\item {29.} M. Jaulent and K. Miodek, Lett. Math. Phys. 1, 243
(1976).\par
\item {30.} A. C. Newell, Solitons in Mathematics and Physics (SIAM,
Philadelphia, 1985).\par
\item {31.} M. Antonowicz and A. P. Fordy, Commun. Math. Phys. 124,
465 (1989). \par
\item {32.}  B.A. Kupershmidt and J. Wilson, Invent. Math. 62, 403
(1981).\par
\item {33.} A.D. Dubrovin, Russian Math. Survey 36, 11 (1981). \par
\item {34.} Yunbo Zeng and Yishen Li, J. Math. Phys. 31, 2835
(1990).\par
\item {35.} J. C. Eilbeck, V. Z. Enol'skii, V. B. Kuznetsov and A. V.
Tsignov, J. Phys. A 27, 567 (1994).\par

\bye

\bye
\bye